\documentstyle{article}
\title{Decoherence in QED at finite temperature}
\author{Z. Haba\\Institute of Theoretical Physics,University of
Wroclaw,Wroclaw,Poland\\e-mail:zhab@ift.uni.wroc.pl}
\date{}
\begin{document}
\maketitle
\begin{abstract}
We consider a  wave packet of a charged particle passing through a cavity
filled with photons at temperature T and investigate
 its localization and  interference properties.
It is shown that the wave packet becomes localized
and the interference disappears
with an exponential speed after a sufficiently long
path through the cavity.

PACS numbers:03.65.-w,03.65.Bz,12.20.Ds
\end{abstract}
\section{Introduction}
It has been known for a long time that thermal photons
can substantially disturb electron beams in an accelerator
\cite{blum}\cite{brown}\cite{deh}. It also has been suggested that
the problem of the classical limit of quantum mechanics
(at least the problem of decoherence) can be
solved in
the presence of a reservoir of photons .
Some  models of an interaction of a quantum system with a reservoir
confirming the decoherent behaviour
have been discussed in refs.\cite{leg}\cite{zur}\cite{bar}.
These models are treated as an approximation to  the quantum electrodynamics
(QED).

We discussed  QED and some approximations to QED in ref.\cite{haba}.
We pointed out a difficulty
 (resulting from the ultraviolet singularity)
 with some approximations to QED when applied in
 a discussion of the decoherence. The singular part of the QED
 propagator does not
 contribute to  the decoherence in formal semiclassical calculations.
In this paper we discuss the regular part.
 We calculate the time evolution of the density matrix of wave packets
in an approximation of a small charge $e$
treating the particle dynamics in a semiclassical approximation.
We neglect the vacuum fluctuations considering the effect of
 thermal photons on the evolution of the density matrix.
 We show a decoherence effect of the thermal photons.
 The vacuum fluctuations lead to
 divergencies at short distances.
 After a  renormalization  the contribution of vacuum fluctuations
 to decoherence is negligible for a large time.

  The effect of photons on decoherence has been discussed first
  by Joos and Zeh \cite{zeh}. The general arrangement for
  interference experiments
  in an environment has been considered in \cite{stern}.
   L. Ford \cite{ford} discussed a change of the vacuum fluctuations
   with a change of the environment's geometry (the Casimir effect) and its
   role in the interference experiments. In such a case
   the relative strength of the vacuum fluctuations
   rather than its intrinsic value (which is infinite)
   is relevant. The  decoherent effect of the black body radiation has been
  studied by Stapp \cite{stapp} but we disagree with
  his results and approximation methods.

We are interested in quantum electrodynamics at finite temperature T
determined by the density matrix (the Gibbs state)
\begin{displaymath}
\rho=\left(Tr\left(\exp\left(-\beta H_{R}\right)\right)\right)^{-1}
\exp(-\beta H_{R})
\end{displaymath}
where $\frac{1}{\beta}=KT$, $K$ is the Boltzmann constant
and $H_{R}$ is the Hamiltonian for the quantum free electromagnetic field.

We  compute the correlation functions explicitely
\begin{equation}
\begin{array}{l}
G_{\beta}(0,{\bf x};t,{\bf x}^{\prime})_{jl} \equiv
Tr\Big(A_{j}(0,{\bf x})A_{l}(t,{\bf x}^{\prime})\rho\Big)=
\cr
 \frac{\hbar c}{\pi^{2}}\int d{\bf k}\vert{\bf k}\vert^{-1}
 \cos\left(\left({\bf x}-{\bf x}^{\prime}\right){\bf k} \right)
 \delta^{tr}_{jl}({\bf k})
\cr
  \Big( cos(c\vert{\bf k}\vert t) \Big(\frac{1}{2} +
   \Big(\exp(\beta\hbar c \vert{\bf k}\vert)-1\Big)^{-1} \Big)
   -\frac{i}{2} sin(c\vert{\bf k}\vert t)\Big)
   \end{array}
\end{equation}
where
\begin{displaymath}
\delta^{tr}_{jl}({\bf k})=\delta_{jl}-k_{j}k_{l}\vert {\bf k}\vert^{-2}
\end{displaymath}

The term $\frac{1}{2}\hbar c\vert{\bf k}\vert$ corresponds to the 
 zero point energy (of vacuum
fluctuations) whereas
$\hbar c \vert{\bf k}\vert (\exp(\hbar c\vert{\bf k}\vert \beta)-1)^{-1}$ 
is the average energy of
thermal photons with the wave number ${\bf k}$.
 The vacuum fluctuation
(noise) is a measurable effect and in general cannot be neglected.
The virtual photons corresponding to the vacuum fluctuations are not
directly observable. Then, the thermal photons are described by the Green's
 function $G_{th}=G_{\beta}-G_{\infty}$
(note that $G_{th}$ is real whereas $G_{\beta}$ and $G_{\infty}$
are complex; in quantum field theory the imaginary part of
the Green's function is related to the pair creation
and annihilation, so subtracting $G_{\infty}$ means that the processes
of pair creation and annihilation are neglected)
\begin{equation}
\begin{array}{l}
 G_{th}({\bf x},{\bf x}^{\prime},t)_{jl}=
 \frac{\hbar c}{2\pi^{2}}\int d{\bf k}\vert{\bf k}\vert^{-1}
 \delta^{tr}_{jl}({\bf k})
 \cos\left(\left({\bf x}-{\bf x}^{\prime}\right){\bf k} \right)
 \cos\left(ct\vert{\bf k}\vert\right)\left(\exp\left(\beta\hbar c
 \vert{\bf k}\vert\right)-1\right)^{-1}
   \end{array}
\end{equation}
Let us note that $G_{th}$ determines the real Gaussian random field.
Subsequent computations can be performed either in the Fock space
or by means of the functional integration . We do not explain the
equivalence of both methods here but refer to our earlier paper
\cite{haba}.

For a small $\vert {\bf x}-{\bf x}^{\prime}\vert$
\begin{equation}
\begin{array}{l}
G_{th}({\bf x},{\bf x}^{\prime},t)_{jl}\simeq G_{th}(0,0,t)_{jl}=\delta_{jl}
\frac{4}{3}\hbar c
 \pi^{-1}
\int_{0}^{\infty} dk k
  cos(ckt)  \Big(\exp(\beta\hbar c k)-1\Big)^{-1}
   \end{array}
\end{equation}
 is approximately ${\bf x}$-independent.

 \section{The semiclassical approximation}

 We  approach the semiclassical limit of the wave function in the
standard way  treating the electromagnetic field ${\bf A}$
as a classical field . Then, the quantum electromagnetic field
is realized as a random field with the covariance (2) .
A solution of the Schr\"odinger equation
\begin{equation}
i\hbar \partial_{t}\psi(t,{\bf x})=\frac{1}{2m}(-i\hbar\nabla +
\frac{e}{c}{\bf A}_{t})^{2}
\psi(t,{\bf x})
\end{equation}
with the initial condition $\psi=exp(\frac{i}{\hbar}W)\phi$
can be related to the solution of the Hamilton-Jacobi equation
 $W_{t}$
\begin{equation}
\partial_{t}W_{t}+\frac{1}{2m}(\nabla W_{t} +\frac{e}{c}{\bf A}_{t})^{2}= 0
\end{equation}
with the initial condition $W_{t=0}({\bf x})=W({\bf x})$.
We express $\psi_{t}$ in the form
\begin{displaymath}
\psi_{t}\equiv\chi_{t}\phi_{t}=exp(\frac{i}{\hbar}W_{t})\phi_{t}
\end{displaymath}
Then, $\psi_{t}$ is the solution of eq.(4) if and only if
$\phi_{t}$ is the solution of the equation
\begin{equation}
\partial_{t}\phi_{t}=\frac{i\hbar}{2m}\triangle \phi_{t}
-\frac{1}{m}(\nabla W_{t}+\frac{e}{c}{\bf A}_{t})\nabla\phi_{t}
-\frac{1}{2m}(\triangle W_{t} + \frac{e}{c}div{\bf A}_{t})\phi_{t}
\end{equation}
with the initial condition $\phi$. In a formal limit
$\hbar \rightarrow 0$ the term $\triangle \phi$ can be
neglected. In such an approximation the solution of the
Schr\"odinger equation (4) is expressed by the classical flow
starting from ${\bf x}$ (here $0\leq s\leq t$)
\begin{equation}
\frac{d{\bf y}_{s}}{ds}=-\frac{1}{m}\big(\nabla W_{t-s}({\bf y}_{s})
+\frac{e}{c}{\bf A}_{t-s}({\bf y}_{s})\big)
\end{equation}
Till $o(\hbar) $ terms we have
\begin{displaymath}
\begin{array}{l}
\psi(t,{\bf x})=exp\Big(\frac{i}{\hbar}W_{t}({\bf x})\Big)
 exp\Bigg(-\int_{0}^{t}\frac{1}{2m}\Big(\triangle W_{t-s}({\bf y}_{s})+
\frac{e}{c}div{\bf A}_{t-s}({\bf y}_{s})\Big)ds\Bigg)
 \phi\big({\bf y}_{t}({\bf x})\big)
\end{array}
\end{displaymath}
If we know the trajectory (e.g., from the Hamilton equations)
then we can compute $W_{t}$
 \begin{equation}
 W_{t}({\bf x})=W({\bf y}_{t}({\bf x}))+\int_{0}^{t}\Big(\frac{m}{2}
 (\frac{d{\bf y}}{ds})^{2} +\frac{e}{c}{\bf A}_{s}({\bf y}_{s})
 \frac{d{\bf y}}{ds}\Big)ds
 \end{equation}
From the correlation functions (2) of ${\bf A}$
it follows that the assumption that ${\bf A}(t,{\bf x})$ is
${\bf x}$-independent is a good approximation for a non-relativistic motion.
For a wave packet of momentum ${\bf p}$ we have $W={\bf px}$. Hence,
approximately
\begin{equation}
\frac{d{\bf y}_{s}}{ds}=-\frac{1}{m}\big({\bf p}+\frac{e}{c}{\bf A}_{t-s}\big)
\end{equation}
This is a consistent approximation because  for  the approximate
solution of the Hamilton-Jacobi
 equation $\nabla W_{s}({\bf x})$ is space independent .
 Then, for  ${\bf A}$ which is space independent we have
 the exact solution of eq.(5)
\begin{equation}
 W_{t}({\bf x})={\bf px}- \frac{t}{2m}{\bf p}^{2}
- \frac{e}{2mc}{\bf p}\int_{0}^{t}{\bf A}_{\tau}d\tau
 -\frac{e^{2}}{2mc^{2}}\int_{0}^{t}{\bf A}_{\tau}^{2}d\tau
 \end{equation}
 As a result of the evolution in an environment of photons
 (which are not under an observation) the
 pure state
 \begin{displaymath}
 \psi=\exp(iW/\hbar)\phi
 \end{displaymath}
 after an average over the states of the quantum electromagnetic field
 is transformed into a mixed state with the density matrix
 \begin{equation}
  \rho_{t}({\bf x},{\bf x}^{\prime})=\langle \overline{\psi}({\bf x})
  \psi({\bf x}^{\prime})\rangle
 \end{equation}
 where the average is over the electromagnetic field.

       We combine the approximation (9)-(10) of a space independent
       ${\bf A}$ with the exact $W_{t}$ (8) in the following
       approximate expression for $W_{t}$
       \begin{equation}
 W_{t}({\bf x})={\bf px}-  \frac{t}{2m}{\bf p}^{2}
- \frac{e}{2mc}{\bf p}\int_{0}^{t}
 {\bf A}_{\tau}({\bf x}-\frac{\tau}{m}{\bf p})d\tau
  \end{equation}
 This expression results from a solution of the equations
 of motion (7) to the lowest (zeroth) order in $e$ . Subsequently,
 $W_{t}$ in eq.(8) is calculated to the first order  in $e$.
 Inserting $W_{t}$ (12) into eq.(11) we obtain
 \begin{equation}
  \begin{array}{l}
  \rho_{t}({\bf x},{\bf x}^{\prime})\equiv
 \exp\Big(
 (i{\bf px}^{\prime}-
  i{\bf px})/\hbar\Big)
  \overline{\phi({\bf x}-\frac{t}{m}{\bf p})}
   \phi({\bf x}^{\prime}-\frac{t}{m}{\bf p})\exp(-S)=
   \cr
 \exp\Big(
 (i{\bf px}^{\prime}-
  i{\bf px})/\hbar\Big)
  \overline{\phi({\bf x}-\frac{t}{m}{\bf p})}
   \phi({\bf x}^{\prime}-\frac{t}{m}{\bf p})
  \cr
  \exp\Big(-\frac{e^{2}}{m^{2}c^{2}\hbar^{2}}\int_{0}^{t}
  {\bf p}G_{th}((s-\tau){\bf p}/m,s-\tau){\bf p}dsd\tau
  \cr
  +\frac{e^{2}}{2m^{2}c^{2}\hbar^{2}}\int_{0}^{t}
  {\bf p}G_{th}({\bf x}-{\bf x}^{\prime}+(s-\tau){\bf p}/m,s-\tau){\bf p}
  dsd\tau
  \cr
+\frac{e^{2}}{2m^{2}c^{2}\hbar^{2}}\int_{0}^{t}
  {\bf p}G_{th}({\bf x}^{\prime}-{\bf x}+(s-\tau){\bf p}/m,s-\tau){\bf p}
  dsd\tau \Big)
\end{array}
  \end{equation}
  where $\exp(-S)$ denotes the last factor in eq.(13).
   If the vacuum fluctuations were to be taken into account
  then we would need to make the replacement $G_{th}\rightarrow
  G_{\beta}=G_{th}+i\triangle_{F}$,
  where  $\triangle_{F}$ is the Feynman causal propagator
  (in the notation of Bjorken and Drell \cite{bjorken}).

\section{The estimates of the evolution of the density matrix}
We perform first the integral over time in eq.(13). We denote
${\bf y}={\bf x}-{\bf x}^{\prime} $, introduce the spherical coordinates
$d{\bf k}=dkk^{2}d\theta\sin\theta d\phi$
and write $S$ in the form
\begin{equation}
 \begin{array}{l}
   S
 =\frac{ e^{2}}{2\pi m^{2}c\hbar}
\int_{0}^{\infty}dk k \int_{0}^{\pi}d\theta \sin\theta I(\theta,k,p)
    \left(\exp\left(\beta\hbar c k\right)-1\right)^{-1}
\end{array}
\end{equation}
We restrict ourselves to $\vert{\bf p}\vert \ll mc$ then
\begin{equation}
\begin{array}{l}
I(\theta,k,p)= c^{-2}k^{-2}
\vert{\bf p}\vert^{2}(1-\cos^{2}\theta)
\cr
\Big(2\left(1-\cos\left( t ck\right)\right)
 - 2\cos{\bf ky}
 +\cos\left( {\bf ky}+ckt\right)
 + \cos\left( {\bf ky}-ckt\right)\Big)
\end{array}
\end{equation}

If ${\bf y}$ is large then on the basis
of eq.(15) the ${\bf y}$-dependent terms are  small
as a function of $t$ in comparison with other terms, because
 we have an additional oscillation in ${\bf y}$ which makes
 such a  term negligible.
 The main contribution to $S$ for a large ${\bf y}$ is
 (we write $\alpha=\cos\theta$)

 \begin{equation}
 \begin{array}{l}
S=\frac{2e^{2}}{m^{2}c^{3}\hbar\pi}{\bf p}^{2}\int_{-1}^{1}d\alpha
(1-\alpha^{2}) \int_{0}^{\infty}dk k^{-1}

\left(1-\cos\left( t ck\right)\right)
    \left(\exp\left(\beta\hbar c k\right)-1\right)^{-1}
\end{array}
\end{equation}
For a small time $t$ ( and a large ${\bf y}$) we obtain
\begin{equation}
\begin{array}{l}
\vert\rho_{t}\vert
\approx
\exp\Big(-\frac{2}{3}\frac{e^{2}}{m^{2}\hbar c\pi}{\bf p}^{2} t^{2}
\int_{0}^{\infty}dk k
    \left(\exp\left(\beta\hbar c k\right)-1\right)^{-1}\Big)
\cr
\approx\exp \left( -
 B\frac{e^{2}}{\hbar c}({\bf p}t/m)^{2} l_{dB}^{-2}
 \right)
 \end{array}
\end{equation}
where $B$ is a constant of order 1 and
   \begin{displaymath}
   l_{dB}=c\hbar \beta
   \end{displaymath}
  denotes the thermal de Broglie wave length at temperature T .
The  result (17) means that the decoherence
(or simply the effect of the electromagnetic  environment) is visible
after a particle makes a path    comparable with de Broglie wave length.

Let us consider now a large $t\geq 0$ in eq.(16). We apply the formula
\begin{displaymath}
1-\cos w=w\int_{0}^{1}d\gamma \sin(\gamma w)
\end{displaymath}
and the formula 3.911 of Gradshtein and Ryzhik  \cite{grad}
\begin{displaymath}
\int_{0}^{\infty}du \sin(au)\Big(\exp(\beta u )-1\Big)^{-1}=
\frac{\pi}{2\beta}\coth(\frac{\pi a}{\beta})-\frac{1}{2a}
\end{displaymath}

Then,
 we obtain
 \begin{equation}
 \begin{array}{l}
S=\frac{4}{3}\frac{te^{2}}{m^{2}c^{2}\hbar\pi}{\bf p}^{2}\int_{0}^{1}d\gamma
 \int_{0}^{\infty}dk
\sin\left( t ck\gamma\right)
    \left(\exp\left(\beta\hbar c k\right)-1\right)^{-1}
\cr
=\frac{4}{3}\frac{e^{2}}{m^{2}c^{3}\hbar\pi}{\bf p}^{2}\int_{0}^{t}d\gamma
\Big(  \frac{\pi}{\beta\hbar } \coth(\frac{\pi\gamma}{\beta\hbar})-
\frac{1}{\gamma} \Big)
\cr
=\frac{4}{3}\frac{e^{2}}{m^{2}c^{3}\hbar\pi}{\bf p}^{2}
\ln\Big(\frac{\beta\hbar}{\pi t}\sinh(\frac{\pi t}{\beta\hbar } )\Big)
 \approx \frac{\vert{\bf p}\vert^{2}}{m^{2}c^{2}}\frac{e^{2}}{c\hbar}
 \frac{ct}{l_{dB}}
\end{array}
\end{equation}
for a large $t$ (and a large $\vert{\bf y}\vert$).

Let us consider next small
$t$ together with a small ${\bf y}$. Then, we can set $s=\tau=0$
 in the argument of $G_{th}$ in eq.(13). In such a case
 \begin{equation}
 \begin{array}{l}
S\simeq\frac{e^{2}}{m^{2}c\hbar\pi}
   {\bf p}^{2} t^{2}
\int_{0}^{\infty}dk k\int_{0}^{\pi}d\theta\sin\theta
 \cr
 \left(1-\cos^{2}\theta\right)
 \left(1-
 \cos \left(k\cos\theta\vert{\bf y} \vert
\right)\right)
    \left(\exp\left(\beta\hbar c k\right)-1\right)^{-1}
   \end{array}
\end{equation}
The formula (19) is relevant only for a small ${\bf y}$ (for a large
${\bf y}$ we return to eq.(17)).
From eq.(19) for a small  ${\bf y}$ and small $t$ we obtain
 \begin{equation}
 \begin{array}{l}
 \rho_{t}({\bf x},{\bf x}^{\prime})\approx
 \exp\Big(-\frac{2e^{2}}{15\pi m^{2}\hbar c} {\bf p}^{2}t^{2}{\bf y}^{2}
 \int_{0}^{\infty}dk k^{3}(\exp(\hbar c\beta k)-1)^{-1}\Big)
\cr
\approx
 \exp\left(-B\frac{e^{2}}{\hbar c}
 \left(\frac{{\bf p}t}{m}\right)^{2}{\bf y}^{2}l_{dB}^{-4}\right)
 \end{array}
 \end{equation}
 where $B$ is a constant of order 1.
 Again we can conclude that the decoherence is visible on distances
 comparable to the thermal de Broglie length.

For a large $t$  we neglect the quickly oscillating
$t$-dependent terms in eq.(15). Then,  similarly as in the computations in
eq.(18) (with $\alpha=cos\theta$)
\begin{equation}
 \begin{array}{l}
S=\frac{e^{2}}{\pi m^{2}c^{3}\hbar}{\bf p}^{2}
\vert{\bf y}\vert\int_{0}^{1}d\gamma\int_{-1}^{1}d\alpha \alpha
(1-\alpha^{2})\int_{0}^{\infty}
dk  \left(\exp\left(\beta\hbar c k\right)-1\right)^{-1}
 \sin\left(k\gamma\alpha\vert{\bf y}\vert\right)
 \cr
=\frac{2e^{2}}{\pi m^{2}c^{3}\hbar}{\bf p}^{2} \int_{0}^{1} d\alpha
(1-\alpha^{2}) \ln\Big(\frac{\beta\hbar c}{\pi \alpha\vert {\bf y}\vert}
 \sinh(\frac{\pi\alpha\vert{\bf y}\vert}{\beta\hbar c})\Big)
 \approx \frac{e^{2}}{\hbar c}\frac{{\bf p}^{2} }{m^{2}c^{2}}
 \frac{\vert{\bf y} \vert}{l_{dB}}
\end{array}
\end{equation}
for a large ${\bf y}$ (and a large $t$).

For a  small $\vert {\bf y}\vert$  (and a large   $t$    )
the  $t$-dependent terms in eq.(15) can be neglected.
Then, the integral (14) gives
\begin{displaymath}
S=\frac{2}{15}\frac{e^{2}}{\pi m^{2}c^{3}\hbar}
\vert {\bf p}\vert^{2}\vert{\bf y}\vert^{2} l_{dB}^{-2}
\int_{0}^{\infty}dk k(\exp k-1)^{-1}
\end{displaymath}
 Hence,
 \begin{displaymath}
\vert \rho_{t}\vert\approx  \exp\Big(-
B\frac{e^{2}}{\hbar c}\vert {\bf p}\vert^{2}(mc)^{-2}
\vert{\bf y}\vert^{2} l_{dB}^{-2}\Big)
\end{displaymath}
At this point it is useful to recall the definition of the Wigner
function
\begin{displaymath}
{\cal W}({\bf q},{\bf k})=(2\pi\hbar)^{-3}\int d{\bf u}\exp(i{\bf ku}/\hbar)
\rho({\bf q}+{\bf u}/2,{\bf q}-{\bf u}/2)
\end{displaymath}
If $\rho\approx \exp(-\frac{a}{2}{\bf y}^{2}-i{\bf py}/\hbar ) $   then
\begin{displaymath}
{\cal W}({\bf q},{\bf k})\simeq \exp(-\frac{1}{2a\hbar^{2}}
({\bf p}-{\bf k})^{2})
w({\bf q},{\bf k})
\end{displaymath}
Such a behaviour means a localization on the classical momentum ${\bf p}$.
 If (as in eq.(18) for a large time) $\rho_{t}=\exp(-i{\bf p}{\bf y}-S)\simeq
 \exp(-i{\bf py}-b{\bf p}^{2}t)$ then approximately
 \begin{equation}
 \partial_{t}\rho_{t}\approx -b[{\bf P},[{\bf P},\rho_{t}]]
 \end{equation}
 where ${\bf P}=-i\hbar\nabla$ is the quantum momentum operator. Then
 \begin{displaymath}
 \partial_{t}{\cal W}_{t}({\bf q},{\bf k})\approx -b{\bf k}^{2}
 {\cal W}_{t} ({\bf q},{\bf k})
 \end{displaymath}

 For N-particles with momenta ${\bf p}_{j}$ we consider
  the initial state of the form
 \begin{displaymath}
 \psi=\exp(\frac{i}{\hbar}\sum_{j=1}^{N}{\bf p}_{j}{\bf x}_{j})\phi
 \end{displaymath}
 Then, $W_{t}$ in eq.(12) is a sum of the Hamilton-Jacobi functions
 for each particle.
 The subsequent expectation value over the electromagnetic field
 gives an exponential with a sum  of pairings between the $N$ particles.
 Hence, it can be expected that $\rho_{t}$ behaves 
 as $\exp(-RN^{2}t^{2})$ for a small t
and as $\exp(-R N^{2}\vert t\vert)$  for a large t with a certain
constant $R$ (note that the behaviour $\exp(-RN\vert t\vert) $ 
for another model has been
 obtained by Unruh \cite{unruh}).
     Such a result is valid for N particles of equal charges
   under the assumption that the terms in the exponential of the form (13)
   have equal signs.
  The contributions add  if the momenta have a distinguished direction.
  If the directions are random then the contributions from different
  particles cancel one another. A similar cancellation takes place
  if the system is neutral , i.e., if the charges $e_{j}$ 
  can have different signs.
  Then, in the sum in an exponential of the form (13)
   we shall have $e_{j}e_{k}{\bf p}_{j}{\bf p}_{k}$  multiplying $G_{th}$.
   Hence,
  there can be cancellations from different charges $e_{j}$ even if
  there is a distinguished direction of the momenta. Under an assumption
   that the contributions
  in the exponential do not cancel one another we obtain    for a large
  distance between any pair of coordinates   and a large $t$
   \begin{displaymath}
   \vert \rho_{t}\vert
 \approx
   \exp\Big(-B\frac{e^{2}}{c\hbar}N^{2}
   (\frac{\vert {\bf p}\vert }{mc})^{2}\frac{ct}{l_{dB}}\Big)
   \end{displaymath}
   where B is a constant of order 1.
   Such a decay rate means that the decay is visible if the distance
   achieved by a particle (with the mean
   momentum ${\bf p}$ ) is comparable
   with the de Broglie wave length. The time needed for this purpose
   can be short
   if $N$ is large
   \begin{displaymath}
   \frac{e^{2}}{c\hbar}N^{2}>(\frac{mc}{\vert 
   {\bf p}\vert })^{2} \frac{l_{dB}}{ct}
   \end{displaymath}
   
 \section{Disappearance of the interference}
 So far we have discussed the decay of $\rho_{t}({\bf x},{\bf x}^{\prime})$
 for varying $t$ and ${\bf y}={\bf x}-{\bf x}^{\prime}$. The disappearance
 of the off-diagonal matrix elements indicates a classical behaviour
 of quantum probabilities in the state $\rho_{t}$.
 We investigate next what happens with the interference describing a typical
 quantum phenomenon.
 We consider a superposition of two
 wave packets
 \begin{displaymath}
 \psi({\bf x})=\exp(i{\bf p}^{(1)}{\bf x}/\hbar)\phi({\bf x})+
 \exp(i{\bf p}^{(2)}{\bf x}/\hbar)\phi({\bf x})
 \end{displaymath}
 Then, at the point ${\bf x}$ (on the screen) after an evolution
 through a cavity filled with thermal photons the probability density
 $\langle\vert \psi_{t}({\bf x})\vert^{2}\rangle $ is
 equal to the diagonal part of the density matrix
 \begin{equation}
 \rho_{t}=\langle\vert \psi_{t}  \rangle  \langle\psi_{t}\vert \rangle
 \end{equation}
  For each packet we have the Hamilton-Jacobi function
  \begin{equation}
  W_{t}({\bf x})={\bf px}-t\frac{{\bf p}^{2}}{2m}-
  \frac{{\bf p}e}{mc}\int_{0}^{t}{\bf A}({\bf y}_{s}({\bf x}),s)ds
  \end{equation}
  Under the time evolution
   \begin{displaymath}
 \psi\rightarrow \psi_{t}=\exp(iW_{t}^{(1)}/\hbar)\phi_{t}^{(1)}+
 \exp(iW_{t}^{(2)}/\hbar)\phi_{t}^{(2)}
 \end{displaymath}
 In our semiclassical  approximation $\phi({\bf x})\rightarrow
\phi({\bf y}_{t}({\bf x}))$). Then, for weak fields
 we neglect the dependence of the paths
  on the electromagnetic field, i.e.,  we  consider straight lines
  \begin{displaymath}
  {\bf y}^{(1)}_{s}={\bf x}-\frac{s}{m}{\bf p}^{(1)}
  \end{displaymath}
  and
  \begin{displaymath}
  {\bf y}^{(2)}_{s}={\bf x}-\frac{s}{m}{\bf p}^{(2)}
  \end{displaymath}
In this approximation the  expectation value (23) is
 \begin{equation}
  \begin{array}{l}
  \langle\mid\psi_{t}({\bf x})\mid^{2}\rangle=
  \vert\phi({\bf x}-\frac{t}{m}{\bf p}^{(1)})\vert^{2}
+  \vert\phi({\bf x}-\frac{t}{m}{\bf p}^{(2)})\vert^{2}     +
\cr
\Big(\overline{\phi({\bf x}-\frac{t}{m}{\bf p}^{(2)})}
   \phi({\bf x}-\frac{t}{m}{\bf p}^{(1)})
  \cr
  \exp\Big(-\frac{i}{\hbar}({\bf p}^{(2)}-{\bf p}^{(1)}){\bf x}
+\frac{it}{2m\hbar}(({\bf p}^{(2)})^{2}-({\bf p}^{(1)})^{2}) \Big) +
 \cr
  \exp\Big(\frac{i}{\hbar}({\bf p}^{(2)}-{\bf p}^{(1)}){\bf x}
-\frac{it}{2m\hbar}(({\bf p}^{(2)})^{2}-({\bf p}^{(1)})^{2}  )\Big)
  \overline{\phi({\bf x}-\frac{t}{m}{\bf p}^{(1)})}
   \phi({\bf x}-\frac{t}{m}{\bf p}^{(2)}) \Big)
  \cr
  \exp\Big(\frac{e^{2}}{2m^{2}c^{2}\hbar^{2}}\int_{0}^{t}
{\bf p}^{(1)}  G_{th}(\frac{s}{m}{\bf p}^{(1)}-\frac{\tau}{m}{\bf p}^{(2)}
,s-\tau)
{\bf p}^{(2)}dsd\tau
\cr
+\frac{e^{2}}{2m^{2}c^{2}\hbar^{2}}\int_{0}^{t}
{\bf p}^{(1)}  G_{th}(\frac{s}{m}{\bf p}^{(2)}-\frac{\tau}{m}{\bf p}^{(1)}
,s-\tau)
{\bf p}^{(2)}dsd\tau
\cr
-\frac{e^{2}}{2m^{2}c^{2}\hbar^{2}}\int_{0}^{t}
  {\bf p}^{(1)}G_{th}(\frac{s}{m}{\bf p}^{(1)}
  -\frac{\tau}{m}{\bf p}^{(1)},s-\tau)
  {\bf p}^{(1)}dsd\tau
  \cr
  -\frac{e^{2}}{2m^{2}c^{2}\hbar^{2}}\int_{0}^{t}
  {\bf p}^{(2)}G_{th}(\frac{s}{m}{\bf p}^{(2)}-\frac{\tau}{m}{\bf p}^{(2)}
  ,s-\tau) {\bf p}^{(2)}
 ds d{\tau} \Big) \equiv \rho_{t}^{(1)}+\rho_{t}^{(2)}+\rho_{t}^{(12)}
\end{array}
  \end{equation}
Without detailed calculations we can obtain the behaviour for a
small t. Let $s=\tau=0$ in $G_{th}$ in eq.(25) then  from eq.(3)
$G_{th}(0,0)\simeq B\delta_{jl}\beta^{-2}\hbar^{-1}c^{-1}$. It follows that
 \begin{equation}
 \begin{array}{l}
 \vert\rho_{t}^{(12)}\vert\simeq
 \exp\Big(-B\frac{e^{2}}{\hbar c}
      l_{dB}^{-2} ( t \vert {\bf p}^{(2)}-{\bf p}^{(1)}\vert/m ) ^{2}
      \Big)
      \end{array}
       \end{equation}
 Hence, if the decoherence is to be visible the distance between the particles
 after time $t$ must be of the order of the de Broglie length.
 The calculations for a large time are more involved.
 Let us denote
 \begin{displaymath}
 \vert\rho_{t}^{(12)}\vert  \equiv \exp(-S_{12})
  \end{displaymath}
   We
   obtain for $S_{12}$
\begin{equation}
  \begin{array}{l}
  S_{12}=\frac{2}{3}\frac{e^{2}}{\pi m^{2}c^{3}\hbar}
    \vert {\bf p}^{(2)}-{\bf p}^{(1)}\vert  ^{2}
  \int_{0}^{\infty}\frac{dk}{k}
  (\exp(\beta\hbar ck)-1)^{-1}
  \left(1-\cos\left(tck \right)\right)
  \cr
  =\frac{2}{3}\frac{e^{2}}{\pi m^{2}c^{3}\hbar}
    \vert {\bf p}^{(2)}-{\bf p}^{(1)}\vert  ^{2}ct
    \int_{0}^{1}d\gamma\int_{0}^{\infty}dk
       (\exp(\beta\hbar ck)-1)^{-1} \sin(\gamma ckt)
       \cr
   =\frac{2}{3}\frac{e^{2}}{\pi m^{2}c^{2}\hbar}
    \vert {\bf p}^{(2)}-{\bf p}^{(1)}\vert  ^{2}
    \int_{0}^{1}d\gamma
       (\frac{t\pi}{2\beta\hbar c}\coth(\frac{\pi\gamma t}{\beta\hbar})-
       \frac{1}{2\gamma c})
       \cr
   
   =\frac{1}{3}\frac{e^{2}}{\pi m^{2}c^{3}\hbar}
    \vert {\bf p}^{(2)}-{\bf p}^{(1)}\vert  ^{2}
    \ln\Big(\frac{\beta\hbar}{\pi t}\sinh
       (\frac{t\pi}{\beta\hbar })\Big)
       \cr
       \approx
    \frac{e^{2}}{\hbar c}\frac{ct }{l_{dB}}
    \vert {\bf p}^{(2)}-{\bf p}^{(1)}\vert  ^{2}m^{-2}c^{-2}
    \end{array}
    \end{equation}
    for a large t.    Hence, the mixed  term in eq.(25) is multiplied
    by $\exp(-S_{12})$ which decays as $\exp(-Rt)$. Such a behaviour
    of the probability density proves that the thermal photons
    lead to the classical  addition of probabilities instead of the
    quantum addition
    of amplitudes showing the decoherence phenomenon in
    a physical model of an electron-photon interaction.

\section{Conclusions}
We have discussed a model of
a quantum charged particle interacting with a quantum
electromagnetic field at finite temperature. We have
calculated the time evolution of the density matrix
in a semiclassical approximation for the wave function and
in a weak coupling approximation for the particle-photon interaction.
In contradistinction to refs.\cite{leg}\cite{zur}\cite{bar}
we do not apply the approximation of a linear  coordinate  coupling
to the environment. For a small time and small space separations
 we obtain an exponential in time
and space
decay $\rho_{t}\approx \exp(-bt^{2}\vert{\bf x}  -{\bf x}^{\prime}\vert^{2})$
of off-diagonal matrix elements (decoherence).
 For a large time the decay achieves its
stationary value (18) and (21). The time and space scale is determined
by the thermal de Broglie wave length. If we have a large number $N$
of charged particles then the decoherence rate can increase as $N^{2}$ .
The density  matrix elements decay as $\rho_{t}\approx \exp(-bt)$
for a large time. Such a behaviour is in agreement with the Lindblad
dynamics if the dissipative part of the dynamics has the form (22).
Such Lindblad dynamics has been discussed in refs.\cite{suta}\cite{zh}.
 The Lindblad dynamics
resulting from a  linear coordinate coupling to the environment
(studied in \cite{leg}\cite{zur}) is
of a different type. It could be described by a replacement of the momentum
operator by a position operator in eq.(22)  (a coupling to the environment
linear in  the momentum as well as in the coordinate has been discussed
 by Leggett in ref.\cite{leggett}).
 We have studied the interference as another typical aspect of
 a quantum behaviour.  We have shown that in an environment
 of photons the interference disappears with an exponential speed
 (eqs.(26)-(27)). Our results suggest an arrangement for
 an experiment. Such experiments could verify the QED beyond
 the usual perturbative approximation as well as  the
 principle of the wave function reduction in a non-selective
 measurement.


\begin{thebibliography}{99}
\bibitem{blum}
 G.R. Blumenthal and R.J. Gould, Rev.Mod.Phys.{\bf 42},237(1970)
 \bibitem{brown}
 L.S. Brown and R. S. Steinke, Am.J.Phys.{\bf 65},304(1997)
 \bibitem{deh}
 B. Dehning et al, Phys.Lett.{\bf B249},145(1990)
\bibitem{leg}
A.O. Caldeira and A.J. Leggett, Physica {\bf A121},587(1983)
\bibitem{zur}
W.G. Unruh and W.H.Zurek, Phys.Rev.{\bf D40},1071(1989)
\bibitem{bar}
P.M.V.B. Barone and A.O.Caldeira,Phys.Rev.{\bf A43},57(1991)
 \bibitem{haba}
Z. Haba, Journ.Math.Phys.{\bf 39},1766(1998)
 \bibitem{zeh}
 E. Joos and H.D. Zeh, Z. Phys. {\bf B59},223(1985)
 \bibitem{stern}
 A. Stern, Y. Aharonov and Y. Imry, Phys.Rev. {\bf A41},3436(1990)
\bibitem{ford}
L.H. Ford, Phys.Rev. {\bf D47},5571(1993)
\bibitem{stapp}
H.P. Stapp, Phys.Rev.{\bf A46},6860(1992)
 \bibitem{bjorken}
J.D. Bjorken and S.D. Drell, 
Relativistic Quantum Fields, McGraw-Hill,1965
 \bibitem{unruh}
 W.G. Unruh, Phys.Rev.{\bf A51},992(1995)
 \bibitem{grad}
I.S. Gradshtein and I.M. Ryzhik, Tables of Integrals,Sums, Series and
Products, Nauka,Moscow,1971 (in Russian)
  \bibitem{suta}
 A. Sandulescu and H. Scutaru, Ann.Phys.(N.Y.) {\bf 173},277(1987)
 \bibitem{zh}
 Z. Haba, Phys.Rev.{\bf A57},4034(1998)
 \bibitem{leggett}
 A.J. Leggett, Phys.Rev.{\bf B30},1208(1984)
 \end{thebibliography}
 \end{document}